\documentclass[12pt]{article}
\usepackage[T2A]{fontenc}
\usepackage[utf8]{inputenc}
\usepackage{euscript}
\usepackage{amssymb}
\usepackage{amsmath}
\usepackage{graphicx}
\usepackage[small,sc,center]{titlesec}
\usepackage{indentfirst}
\usepackage{siunitx}

\newcommand{\nc}{\newcommand}
\nc{\aov}{\alpha_\mathrm{ov}}
\nc{\K}{\:\mathrm{K}}
\nc{\mzams}{M_\mathrm{ZAMS}}
\nc{\teff}{T_\mathrm{eff}}
\nc{\tev}{t_\mathrm{ev}}

\begin{document}

\begin{center}
\textbf{Evolution, pulsation and period change in the Cepheid SZ Tau}

\vskip 3mm
\textbf{Yu. A. Fadeyev\footnote{E--mail: fadeyev@inasan.ru}}

\textit{Institute of Astronomy, Russian Academy of Sciences, Pyatnitskaya ul. 48, Moscow, 119017 Russia} \\

Received June 17, 2015
\end{center}

\textbf{Abstract} ---
The study is devoted to radial pulsations in Population~I Cepheids with masses
from $5.4M_\odot$ to $6M_\odot$.
Solution of the equations of radiation hydrodynamics and turbulent convection
for nonlinear stellar pulsations was obtained with initial conditions as
the core helium burning models of computed evolutionary sequences.
For each value of the initial mass we considered stellar evolution
from the zero age main sequence to central helium exhaustion with
three values of the convective overshoot parameter: $\aov=0.1$, 0.15 and 0.2.
Models for the Cepheid SZ~Tau with pulsation period $\Pi=3.149$~day
can be constructed only for $\aov=0.1$ and $\aov=0.15$.
The star is at the evolutinary stage of the second crossing of the instability
strip and pulsates in the first overtone just near the boundary between
the fundamental mode and the first overtone.
The mass, radius and age of the star are in the ranges
$5.46\le M/M_\odot\le 5.75$, $41.5\le R/R_\odot\le 42.3$ and
$6.9\times 10^7~\mathrm{yr} \le\tev\le 8.0\times 10^7~\mathrm{yr}$,
respectively.
Predicted period change rates are of $\dot\Pi\approx -0.4$~s/yr.


\newpage
\section*{introduction}

The pulsating variable star SZ~Tau belongs to a small group of Cepheids that are
members of open clusters.
The first report that the Cepheid SZ~Tau observed in the extended corona of NGC~1647
belongs to this cluster was made by Efremov (1964) on the basis of similar values
of the distance and the radial velocity of the Cepheid and the stars of the cluster.
Later this conclusion was confirmed by other observational data
(Gieren 1985; Turner 1992; Anderson et al. 2013).
The study of the Cepheid which belongs to the stellar cluster and determination of its
fundamental parameters (the mass, radius and luminosity) by methods of
stellar pulsation theory is of great importance since provides us with
an independent method of evaluation of the distance and the age of the stellar cluster.

In the General Catalogue of Variable Stars (Samus et al. 2012) the star SZ~Tau
with a period $\Pi=3.14873$~day is referred to Cepheids of type DCEPS
with nearly symmetrical light curves of small amplitude.
Turner (1992) estimated that the average absolute magnitude of SZ~Tau
corresponds to radial pulsations in the first overtone.
Later this assumption was confirmed by Fourier analysis of the radial velocity
curve (Antonello and Aikawa 1995).
Another confirmation of the first overtone pulsation was provided in the
work by Bono et al. (2001) where the mean stellar radius $R=45.6R_\odot$
obtained from the radial velocity observations (Bersier et al. 1994)
was compared with empirical period--radius relation of Calactic Cepheids.
It should be noted that the mean radius of SZ~Tau was evaluated by the
Baade--Wesselink method in a number of studies
(Gieren 1985; Moffett and  Barnes 1987; Laney and Stobie 1995;
Ripepi et al. 1997; Rastorguev and Dambis 2011)
but the large spread in values of the radius
($34R_\odot\le R \le 57R_\odot$)
is a serious handicap for determining the stellar mass from the period--mean
density relation.

In contrast to most of the Cepheids the light curve of SZ~Tau
reveals the absence of strictly repetitive light variability and this
feature substantially complicates the detection of period change due
to stellar evolution.
In particular, Szabados (1977) and Evans et al. (2015) found
significant erratic variations in the $O-C$ diagram
which do not allow for an unambiguous detection of the secular
period change.
However Berdnikov and Pastukhova (1995) showed that the $O-C$ diagram
can be fitted by the quadratic function and therefore SZ~Tau is on the
evolutionary stage of the second crossing of the Cepheid instability strip
and its period decreases with a rate of $\dot\Pi=-0.49$~s/yr.

The goal of the present study is to theoretically estimate the fundamental
parameters of SZ~Tau and to validate the conclusion of
Berdnikov and Pastukhova (1995) on the evolutionary status of the star.
To this end we carry out the consistent calculations of stellar evolution and
nonlinear stellar pulsations and determine the conditions for pulsations
with period of $\Pi=3.149$~day, whereas the theoretical estimates of the period
change rate are compared with $\dot\Pi=-0.49$~s/yr.
In this work we do not use the period--mean density relation and therefore
evaluation of the stellar mass does not suffer from uncertainties in the stellar
radius estimates obtained by the Baade--Wesselink method.
Earlier the author employed such an approach for determination of the
fundamental parameters and the age of the Cepheid $\alpha$~UMi (Fadeyev 2015).

Basic equations and methods of their solution are described in previous
papers of the author (Fadeyev 2013a, 2014, 2015).
Results of computations presented below are obtained for the initial
composition $X=0.7$, $Z=0.02$, where $X$ and $Z$ are the mass fractional
abundances of hydrogen and of elements heavier than helium, respectively.
Convection in evolving stars is treated according to the mixing length theory
(B\"ohm--Vitense 1958) with a mixing length to pressure scale height ratio
$\alpha_\Lambda=1.6$.
Enlargement of the convective core due to convective overshooting is taken
into account by the parameter $\aov = \Delta r/H_\mathrm{P}$, where
$\Delta r$ is the radial extention of the outer boundary of the convective core
and $H_\mathrm{P}$ is the pressure scale height.
The evolutionary computations were done for $\aov=0.1$, 0.15, 0.2 and
the parameter $\aov$ was assumed to be the same from the main sequence to
core helium exhaustion.

\section*{results of computations}

To determine the ranges of the initial stellar mass $\mzams$ and the parameter $\aov$
for which the radial pulsations with a period of $\Pi=3.149$~day become possible
during the second crossing of the instability strip we computed the grid of
evolutioinary tracks for stars with masses $5.4M_\odot\le\mzams\le 6M_\odot$.
Together with computation of stellar evolution we solved the linear adiabatic
wave equation (see, e.g., Cox 1983) and evaluated eigenfrequencies of the
fundamental mode and the first overtone.
If the evolutionary model with effective temperature $5400\K\le\teff\le 6700\K$
has the adiabatic period of one of these oscillation modes close to the value
$\Pi=3.149$~day then from 10 to 15 models of the evolutionary sequence were
selected and were used as initial conditions for solution of the equations of
hydrodynamics.

From results of hydrodynamic computations we primarily determined the edges
of the pulsation instability strip, that is the age of the star $\tev$
corresponding to the zero growth rate of the kinetic energy.
Subsequently for each hydrodynamic model we calculated the periods of the
fundamental mode and the first overtone using the Fourier transform of the
kinetic energy of pulsation motions.
In the most of considered evolutionary sequences the mode switch from the
fundamental mode to the first overtone occurs while the star crosses
the instability strip.
In such a case it was assumed that the mode switch is abrupt and the change
of the pulsation period $\Pi$ within the instability strip is described
by two continuous functions.
Determination of the mode switch boundary is discussed in our previous papers
(Fadeyev, 2013b; 2015).
Within the evolutionary time interval with continuous change of the period
the sequence of the values of $\Pi$ obtained from hydrodynamic computations
is fitted by the algebraic polynomial of the order $2\le n\le 4$.
The choice of the polynomial order $n$ obeys the condition that the
approximation rms error cannot exceed 0.1\%.

For Cepheid hydrodynamic models computed in the present study
the most short period of the fundamental mode is $\Pi\approx 4.3$~day
therefore the following discussion of SZ~Tau model will be confined
to consideration of the first overtone pulsators.
The existence of the point of the evolutionary track with
pulsation period $\Pi=3.149$~day
depends on both the initial stellar mass $\mzams$ and the overshoot parameter
$\aov$.
Dependence on the initial mass is illustrated in Fig.~\ref{fig1} where
the evolutionary tracks in the Hertzsprung--Russel (HR) diagram are
shown for the models $\mzams=5.5M_\odot$ and $\mzams=5.8M_\odot$
computed with the overshoot parameter $\aov=0.15$.
For the sake of graphical clarity, we present in the figure only
the blue loops corresponding to the core helium burning.
Evolution along the shown tracks proceeds clockwise.

As soon as a star with an initial mass $\mzams=5.5M_\odot$ crosses the red edge
of the instability strip ($\teff=5550\K$) it becomes a fundamental mode pulsator.
The pulsation switches to the first overtone at the effective temperature
$\teff=5840\K$ and the pulsation ceases at the blue edge of the instability
strip ($\teff=6300\K$).
As can be seen in Fig.~\ref{fig1}, the model $\mzams=5.5M_\odot$ with pulsation
period $\Pi=3.149$~day locates just near the boundary between the fundamental
mode and the first overtone,

As the initial stellar mass increases the boundary of the mode switch from
the fundamental mode to the first overtone moves to higher values of $\teff$
 and for stars $\mzams\ge 5.8M_\odot$ becomes beyond the blue edge of the
instability strip.
Therefore, as seen in Fig.~\ref{fig1}, the star with initial mass
$\mzams=5.8M_\odot$ remains the fundamental mode pulsator within the whole
instability strip ($5360\K\le\teff\le6250\K$).
The pulsation periods are confined to the interval
$4.45~\mathrm{day}\le\Pi\le 6.90~\mathrm{day}$.

Conditions for the occurence of pulsations with period $\Pi=3.149$~day
on the evolutionary tracks computed with different values of the
overshoot parameter $\aov$ are illustrated in Fig.~\ref{fig2}
for $\mzams=5.6M_\odot$.
As can be seen, the luminosity of the Cepheid increases with increasing
$\aov$ (mainly due to the larger mass of the convective core during the
main sequence stage), whereas the maximum effective temperature of the
blue loop decreases.
As a result, among three evolutionary tracks shown in Fig.~\ref{fig2}
only one ($\aov=0.15$) can be used for construction of the SZ~Tau
model, the star locating just near the boundary between the fundamental
mode and the first overtone.
The evolutionary track computed with $\aov=0.1$ is not appropriate for
the model of SZ~Tau because of too short period ($\Pi=2.81$~day)
of the first overtone at the mode switch boundary.
On the other hand, for the evolutionary track computed with $\aov=0.2$
the blue edge of the instability strip is beyond the turning point of the track
so that during the stage of increasing effective temperature the Cepheid
remains the fundamental mode pulsator.

It should be noted that for all evolutionary tracks
($5.4M_\odot < \mzams < 6M_\odot$) computed with $\aov=0.2$
the radial pulsations during the second crossing of the instability
strip were found to be due to the instability in the fundamental mode.
Therefore further discussion will be confined to results of hydrodynamical
computations done with initial conditions obtained with
$\aov=0.1$ and $\aov=0.15$.

Conditions of existence of the pulsation with period $\Pi=3.149$~day
can be represented in more detail by the dependence of the period of the
first overtone on the evolutionary time.
The plots for several evolutionary sequences computed with $\aov=0.1$
are shown in Fig.~\ref{fig3}.
Each plot ends at the blue edge of the instability strip.
For the sake of graphical convenience the evolutionary time $\tev$ is
set to zero at the mode switch boundary.

For $\mzams\ge 5.9M_\odot$ the star pulsates in the fundamental mode
within the whole instability strip.
On the other hand, for $\mzams\le 5.6M_\odot$ the period of the first
overtone is smaller than $\Pi=3.149$~day.
Therefore, the model of SZ~Tau with the overshoot parameter $\aov=0.1$
can be constructed for the initial stellar mass $\mzams$ ranging from
$5.6M_\odot$ to $5.9M_\odot$.
The principal cause of the narrow mass interval is the close position
of the Cepheid to the mode switch boundary.

Temporal dependences of the first overtone period in the evolutionary
sequences computed with overshoot parameter $\aov=0.15$ show
a similar behavior.
The only difference in comparison with models
computed with $\aov=0.1$ is that
pulsations with period $\Pi=3.149$ day arise in stars with lower
masses: $\mzams=5.5M_\odot$ and $\mzams=5.6M_\odot$.
The narrow ranges of initial stellar masses allowed us to approximately
evaluate the fundamental parameters of SZ~Tau which are given
in the table.

To compare our theoretical results with observations Fig.~\ref{fig4}
shows the plots of $\dot\Pi$ as a function of period $\Pi$
for Cepheid models pulsating in the first overtone.
Each curve in Fig.~\ref{fig4} represents the evolutionary track in the
period--period change rate diagram and evolution proceeds as the
pulsation period $\Pi$ decreases.
The position of SZ~Tau on the diagram and its rms error are shown
on the basis of data given by Berdnikov and Pastukhova (1995).

\subsection*{CONCLUSIONS}

The Cepheid SZ~Tau is located just near the mode switch from
the fundamental mode to the first overtone.
The proximity of the pulsation period $\Pi=3.149$~day to the
boundary value of the period of first overtone is
by an order of magnitude $\delta\Pi/\Pi\sim 10^{-2}$.
This is the principal cause that stellar evolution and
nonlinear stellar pulsation calculations lead to the 
narrow ranges of the fundamental parameters of the Cepheid.
It should also be noted that models of SZ~Tau
obtained for the overshoot parameters $\aov=0.1$ and $\aov=0.15$
give evidence in favor of the small and intermediate convective
overshooting during the main sequence and the helium core burning
stages.

Predicted period change rates obtained in the present study
are in a good agreement with the observational estimate
($\dot\Pi=-4.9$~s/yr) by Berdnikov and Pastukhova (1995)
and therefore confirm a suggestion that SZ~Tau is at the
evolutionary stage of the second crossing of the
instability strip.

Theoretical estimates of the radius of SZ~Tau
($41.5R_\odot\le R\le 42.3R_\odot$) given in the table
show better agreement with measurements by Ripepi et al. (1997).
Authors of this work employed the improved version of the
Baade--Wesselink method (the CORS method) based on the more
correct relations between the color index and the surface
brightness (Caccin et al. 1981).
For two variants of the CORS method Ripepi et al. (1997)
estimated the mean radius as $R=41.8R_\odot$ and $R=44.8R_\odot$.

Hydrodynamic computations of nonlinear pulsations in SZ~Tau
located just at the mode switch are accompanied with great
numerical difficulties.
This is mainly due to the small decay rate of the
fundamental mode which leads to slow relaxation of
nonregular oscillations after attainment of the first overtone
limit cycle.
Absence of strict regularity in the observed light and radial
velocity curves of SZ~Tau might be due to slow decay of the
fundamental mode.

\newpage
\subsection*{REFERENCES}
\begin{enumerate}

\item R. I. Anderson, L. Eyer, and N. Mowlavi, MNRAS \textbf{434}, 2238 (2013).

\item E. Antonello and T. Aikawa, Astron. Astrophys. \textbf{302}, 105 (1995).
  
\item L. N. Berdnikov and E. N. Pastukhova, Pis'ma Astron. Zh. \textbf{21}, 417 (1995)
      [Astron. Lett. \textbf{21}, 369 (1995)].

\item D. Bersier, G. Burki, M. Mayor, et al., Astron. Astrophys. Supppl. \textbf{108}, 25 (1994).

\item E. B\"ohm--Vitense, Zeitschrift f\"ur Astrophys. \textbf{46}, 108 (1958).

\item G. Bono, W. P. Gieren, М. Marconi, et al., Astrophys. J. \textbf{552}, L141 (2001).

\item R. Caccin, A. Onnembo, G. Russo, et al., Astron. Astrophys. \textbf{97}, 104 (1981).

\item J. P. Cox, \textit{The Theory of Stellar Pulsation} (Princeton Univ., Princeton, 1980; Mir, Moscow, 1983).

\item Yu. N. Efremov, Perem. Zv. \textbf{15}, 242 (1964).

\item N. R. Evans, R. Szab\'o, A. Derekas, et al., Mon. Not. R. Astron. Soc. \textbf{446}, 4008 (2015).

\item W. P. Gieren, Astron. Astrophys. \textbf{148}, 138 (1985).

\item C. D. Laney and R. S. Stobie, Mon. Not. R. Astron. Soc. \textbf{274}, 337 (1995).

\item T. J. Moffett and T. J. Barnes III, Astrophys. J. \textbf{323}, 280 (1987).

\item A. S. Rastorguev and A. K. Dambis, 2011, Astrophys. Bull. \textbf{66}, 47 (2011).

\item V. Ripepi, F. Barone, L. Milano, et al., Astron. Astrophys \textbf{318}, 797 (1997).

\item N. N. Samus, O. V. Durlevich, E. V. Kazarovets, et al., General Catalogue of Variable Stars (GCVS database, version April 2012), CDS B/gcvs (2012).

\item L. Szabados, Commun. Konkoly Observ. \textbf{70}, 1  (1977).

\item D. G. Turner, Astron. J. \textbf{104}, 1865 (1992).

\item Yu. A. Fadeyev, Pis'ma Astron. Zh. \textbf{39}, 342 (2013a)  
      [Astron.Lett. \textbf{39}, 306 (2013a)].

\item Yu. A. Fadeyev, Pis'ma Astron. Zh. \textbf{39}, 829 (2013b)  
      [Astron.Lett. \textbf{39}, 746 (2013b)].

\item Yu. A. Fadeyev, Pis'ma Astron. Zh. \textbf{40}, 301 (2014)  
      [Astron.Lett. \textbf{40}, 301 (2014)].

\item Yu. A. Fadeyev, Mon. Not. R. Astron. Soc. \textbf{449}, 1011 (2015).

\end{enumerate}

\newpage
\begin{center}
Models of Cepheids with period 3.149 day

\begin{tabular}{l|S|S|r|r|r|r|r}
\hline
 $\aov$ & $\mzams/M_\odot$ & $M/M_\odot$ & $R/R_\odot$ & $L/L_\odot$ & $\teff$, K  & $\dot\Pi$, s/yr & $\tev, 10^6$ yr \\
\hline
 0.1  & 5.7 & 5.66 & 42.0 & 1878 & 5870 & -0.42 & 72.1 \\
      & 5.8 & 5.75 & 42.3 & 2008 & 5950 & -0.43 & 69.2 \\
 0.15 & 5.5 & 5.46 & 41.5 & 1819 & 5860 & -0.38 & 80.2 \\
      & 5.6 & 5.56 & 41.8 & 1968 & 5960 & -0.43 & 77.0 \\
\hline
\end{tabular}
\end{center}
\clearpage

\newpage
\section*{Figure captions}

\begin{figure}
\centerline{\includegraphics[width=15cm]{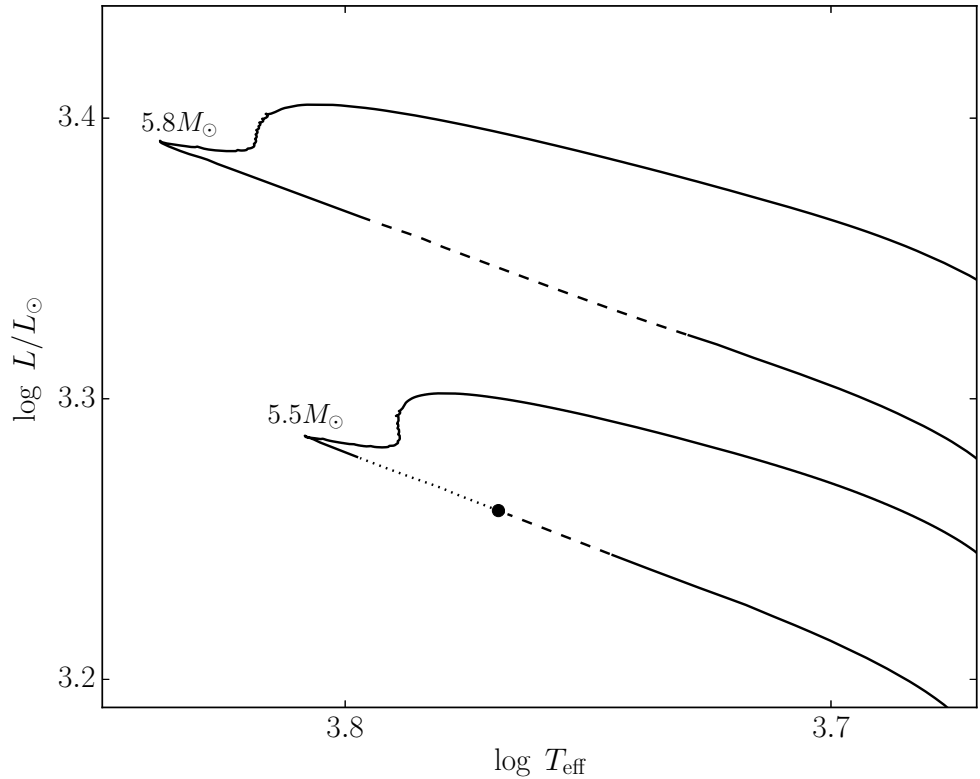}}
\caption{Evolutionary tracks of core helium burning stars with initial masses
         $5.5M_\odot$ and $5.8M_\odot$ computed with the overshoot parameter $\aov=0.15$.
         Parts of the track corresponding to the pulsational instability during
         the second crossing of the instability strip are shown by the dashed
         and dotted lines for the fundamental mode and the first overtone, respectively.
         The filled circle indicates the position of the Cepheid pulsating
         in the first overtone with period $\Pi=3.149$~day.}
\label{fig1}
\end{figure}

\begin{figure}
\centerline{\includegraphics[width=15cm]{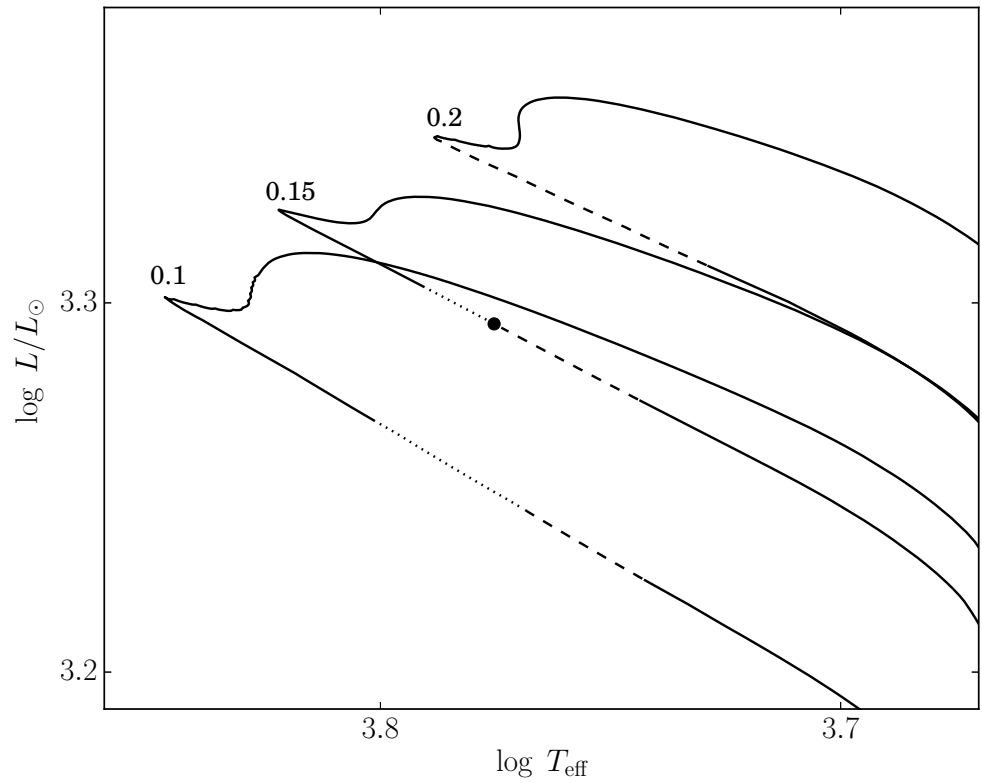}}
\caption{Same as Fig.~\ref{fig1} but for the initial stellar mass $\mzams=5.6M_\odot$
         with $\aov=0.1$, 0.15 and 0.2.}
\label{fig2}
\end{figure}

\begin{figure}
\centerline{\includegraphics[width=15cm]{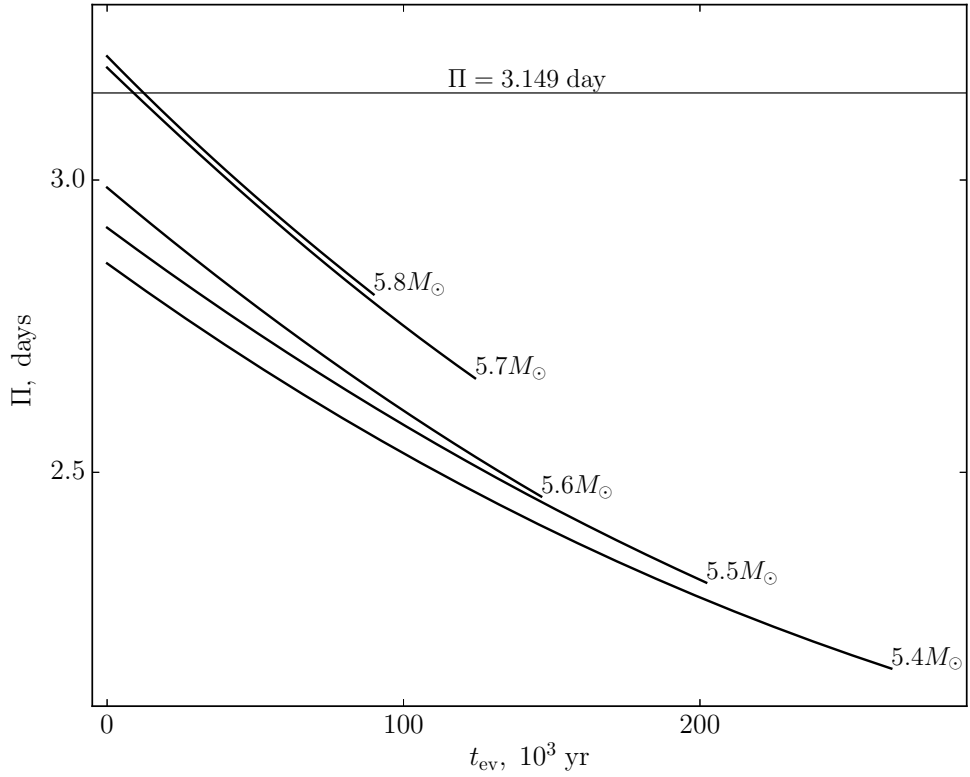}}
\caption{The first overtone pulsation period $\Pi$ as a function of the evolutionary time
         $\tev$ during the second crossing of the instability strip.
         Solid lines show evolutionary sequences computed for $\aov=0.1$.
         The evolutionary time $\tev=0$ corresponds to the mode switch from
         the fundamental mode to the first overtone.}
\label{fig3}
\end{figure}

\begin{figure}
\centerline{\includegraphics[width=15cm]{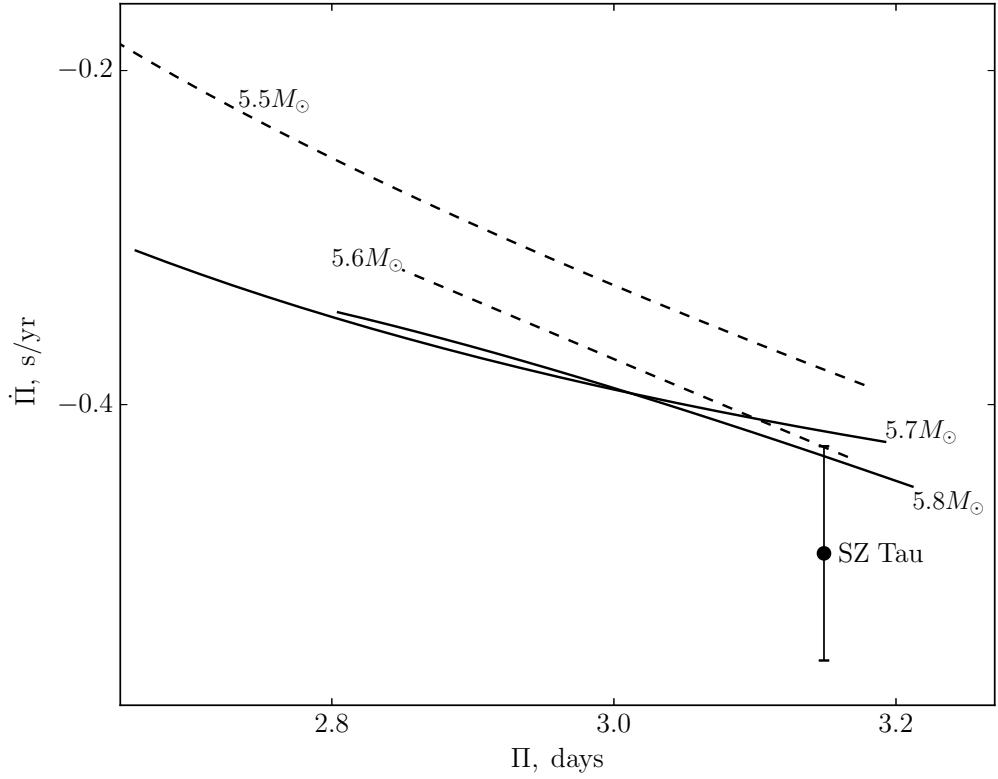}}
\caption{The period change rate $\dot\Pi$ as a function of period $\Pi$
         for Cepheids pulsating in the first overtone.
         The solid and dashed lines show the evolutionary sequences
         computed for the overshoot parameter $\aov=0.1$ and $\aov=0.15$,
         respectively.
         The values of $\mzams$ are given at the curves.
         The position of the Cepheid SZ~Tau corresponds to the
         observational estimates by Berdnikov and Pastukhova (1995).}
\label{fig4}
\end{figure}

\begin{itemize}
\item[Fig. 1.] 
         Evolutionary tracks of core helium burning stars with initial masses
         $5.5M_\odot$ and $5.8M_\odot$ computed with the overshoot parameter $\aov=0.15$.
         Parts of the track corresponding to the pulsational instability during
         the second crossing of the instability strip are shown by the dashed
         and dotted lines for the fundamental mode and the first overtone, respectively.
         The filled circle indicates the position of the Cepheid pulsating
         in the first overtone with period $\Pi=3.149$~day.

\item[Fig. 2.]
         Same as Fig.~\ref{fig1} but for the initial stellar mass $\mzams=5.6M_\odot$
         with $\aov=0.1$, 0.15 and 0.2.

\item[Fig. 3.]
         The first overtone pulsation period $\Pi$ as a function of the evolutionary time
         $\tev$ during the second crossing of the instability strip.
         Solid lines show evolutionary sequences computed for $\aov=0.1$.
         The evolutionary time $\tev=0$ corresponds to the mode switch from
         the fundamental mode to the first overtone.

\item[Fig. 4.]
         The period change rate $\dot\Pi$ as a function of period $\Pi$
         for Cepheids pulsating in the first overtone.
         The solid and dashed lines show the evolutionary sequences
         computed for the overshoot parameter $\aov=0.1$ and $\aov=0.15$,
         respectively.
         The values of $\mzams$ are given at the curves.
         The position of the Cepheid SZ~Tau corresponds to the
         observational estimates by Berdnikov and Pastukhova (1995).
\end{itemize}

\newpage

\newpage

\newpage

\newpage

\end{document}